\documentclass[a4paper,fleqn,usenatbib]{mnras}

\usepackage[T1]{fontenc}
\usepackage{ae,aecompl}

\usepackage{graphicx}	
\usepackage{amsmath}	
\usepackage{amssymb}	
\usepackage{lipsum}

\title[Pulse Nulling in PSR B1706$-$16]{Detection of long nulls in PSR B1706$-$16, a pulsar with large timing irregularities.}

\author[A. Naidu et al.]{
Arun Naidu,$^{1,2}$\thanks{E-mail: arun@ncra.tifr.res.in}
Bhal Chandra Joshi,$^{1}$\thanks{E-mail: bcj@ncra.tifr.res.in}
P.K Manoharan$^{1,3}$
and M.A Krishnakumar$^{1,3}$
\\
$^{1}$National Centre for Radio Astrophysics, Pune, India\\
$^{2}$McGill Space Institute, McGill University, Montreal, Quebec, Canada\\
$^{3}$Radio Astronomy Centre , Udhagmandalam, Tamilnadu, India\\
}

\date{Accepted for publication in MNRAS.}

\pubyear{2016}

\begin{document}
\label{firstpage}
\pagerange{\pageref{firstpage}--\pageref{lastpage}}
\maketitle

\begin{abstract}
Single pulse observations, characterizing in detail, the nulling 
behaviour of PSR B1706$-$16 are being reported for the first time 
in this paper. Our regular long duration monitoring of this 
pulsar reveals long nulls of 2 to 5 hours with an overall 
nulling fraction of 31$\pm$2\%. The pulsar shows two distinct 
phases of emission. It is usually in an active phase, characterized 
by pulsations interspersed with shorter nulls, with a nulling fraction 
of about 15 \%, but it also rarely 
switches to an inactive phase, consisting of long nulls. The nulls in 
this pulsar are concurrent between 326.5 and 610 MHz. Profile 
mode changes accompanied by changes in fluctuation properties 
are seen in this pulsar, which switches from mode A before a null 
to mode B after the null. The distribution of null durations in 
this pulsar is bimodal.  With its occasional long nulls, PSR 
B1706$-$16 joins the small group of intermediate nullers, 
which lie between the classical nullers and the intermittent 
pulsars. Similar to other intermediate nullers, PSR B1706$-$16 
shows high timing noise, which could be due to its rare long nulls 
if one assumes that the slowdown rate during such nulls 
is different from that during the bursts.
\end{abstract}




\section{Introduction}
Pulsars are highly magnetized rotating neutron stars, 
which emit coherent beamed electromagnetic emission 
at the expense of their rotational energy, 
effectively slowing down the star. Their pulsed emission 
varies from pulse to pulse. Absence of this pulsed emission 
for several pulsar rotations was first noted by \cite{Backer1970} 
in four pulsars. This phenomenon, called pulse nulling, has since 
been seen in more than 100 pulsars to date 
\citep{Wang2007,Biggs1992,Ritchings1976,Burke2012,gjk12,gjw14}. 
The duration of nulls varies not only from one pulsar to other, 
but also for a given pulsar. The percentage of pulses without 
detectable emission is called nulling fraction (NF), which 
ranges from few percent to more than 90 percent. While pulsars such as 
PSR B0835$-$41 and B2021+51 show mostly single pulse nulls 
\citep{gjk12}, no emission is seen in PSR B0826$-$34 for 
15000 pulses \citep{dll+79}. Previously discovered intermittent 
pulsars such as PSR B1931+24 \citep{klo+06}, PSR J1841$-$0500 
\citep{crc+12}, PSR J1832$+$0029 \citep{llm+12}, PSR J1107$-$5907 \citep{yws+14},
PSR J1910+0517 and J1929+1357 \citep{lsf+16}, where no pulsed emission 
is observed from few days to several years, can also 
be considered as neutron stars with an extreme form of nulling. 
Interestingly, the rate of slowdown ($\dot{\nu}$) is reduced 
in these intermittent pulsars during their inactive phase 
suggesting changes in torque \citep{klo+06,Lyne2009}. Changes 
in magnetosphere state were proposed to explain the inactive phase 
in these pulsars as this steers the emission beam away 
from the line of site  in addition to a change in slowdown 
rate \citep{Timokhin2010}. Hence, some form of rotation rate 
irregularities are expected in pulsars, which fall 
in between classical nullers and intermittent pulsars.

In recent years, there is a growing class of such intermediate 
nullers with nulling time scales of a few hours. Good examples 
are PSRs  like B0823+26 \citep{Young2012}, 
PSR J1717$-$4054 \citep{Johnston1992}, PSR J1634$-$5107 
\citep{OBrien2006} and  PSR J1853+0505 \citep{Young2015}. 
Frequent emission of 
long nulls results in high NFs ($> 70\%$) for all these 
pulsars. Unlike the intermittent pulsars, where the change 
in $\dot{\nu}$ can be estimated during the absence of emission 
for several days, it is difficult to estimate the slowdown 
rates for intermediate nullers as the duration of null phase 
is not long enough to see a significant difference through 
pulsar timing.

In this paper, detailed single pulse observations of  
PSR B1706$-$16 (PSR J1709$-$1640), discovered in one of the initial 
Molonglo {\bf surveys} \citep{Large1969}, are presented. It is like 
any other normal pulsar with a period of 653 ms and 
dispersion measure (DM) of 24.8733 pc/cm$^3$ (Table \ref{psrparam}). While no nulling 
was reported even 4 decades after its discovery, 
it was identified as nuller in a single pulse follow 
up study of High Time Resolution Universe 
survey \citep{Burke2012}. Its single pulse studies 
are also relatively undocumented. The pulsar also shows 
an interesting red noise distribution of the timing 
residuals as reported by \cite{Baykal1999}. In our 
study, PSR B1706$-$16  has shown long nulls ($>$ 2 hrs) 
about once in a week making this a unique addition 
to the class of intermediate nullers. In Section 
\ref{observations}, the observations are described.  
The description of analysis and results is in 
Section \ref{analysis} followed by discussions  
and conclusions in Section \ref{discussion} and 
Section \ref{conclusion} respectively. 

\begin{table*}
 \caption{The known parameters for PSR B1706$-$16}
 \label{psrparam}
 \centering
 \begin{tabular}{|l|c|c|c|c|c|c|c}
  \hline
  JNAME     & Right Ascension & Declination &Period  &  DM  & S$_{400 MHz}$ & Surface Magnetic field & Characteristic Age \\
            & (h:m:s) & (d:m:s) &(s)     &(pc\,cm$^{-3}$)& (mJy ) & (10$^{12}$ G) & (Myr) \\
  \hline
J1709$-$1640&17:09:26.44&-16:40:57.73&0.653054&24.89&47&2.05 &  1.64 \\
  \hline
 \end{tabular}
\end{table*}

\begin{figure}
	
	\includegraphics[width=1.0\columnwidth]{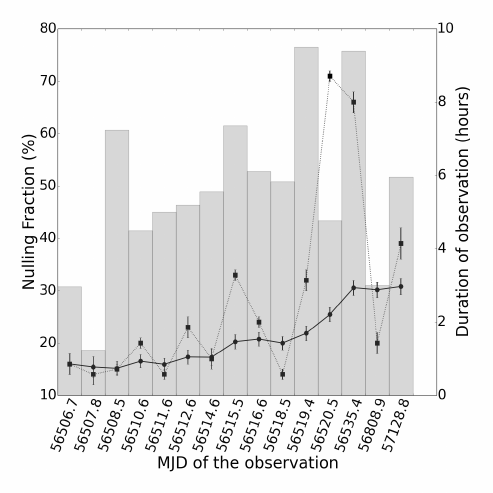}
	\centering
    \caption{Histogram showing the date of observations, 
duration of observations (bars in the histogram) and 
corresponding nulling fractions. The dotted line represents 
the nulling fraction at each epoch and the solid line 
represents the cumulative nulling fraction with 
error bars. The ordinate on the left represents {\bf the} 
nulling fraction and that on the right represents 
duration of observations at each epoch.}
    \label{nulling}
\end{figure}


\section{Observations}
\label{observations}

All the 325 MHz observations were carried out using the 
Ooty Radio Telescope [ORT,\cite{Swarup1971}], which is 
a single dish cylindrical parabolic reflector with 
linearly polarized dipole feed. Data were recorded 
by using the newly commissioned pulsar receiver 
PONDER \citep{Naidu2015}. The frequency of observations 
was 326.5 MHz with a bandwidth of 16 MHz. 
The ORT has a capability to track PSR 
B1706$-$16 for about 9.5 hours. A total of 15 long 
duration observations were carried out for this 
pulsar. The duration of observations varied from 
2 hours to 9.5 hours (see Figure \ref{nulling}). 
All the data were recorded after incoherent 
dedispersion at the pulsar's nominal  DM of 
24.873 pc/cm$^{-3}$ and timeseries was provided 
in SIGPROC\footnote{http://sigproc.sourceforge.net} 
format. Further, daily short observations were 
carried out over a period of four months to 
check for any timing irregularities and to obtain 
an updated timing solutions for the analysis. 

In addition, simultaneous multi-frequency 
observations were carried out using the 
Giant Meterwave Radio Telescope (GMRT) \citep[]{sak+91} 
and the ORT for this pulsar. The data at 
the GMRT were obtained using the GMRT 
Software Backend (GSB) \citep{Roy2010} in a 
phased array mode, where closely spaced 15 
antennas were phased to form a single beam 
in the sky. The GMRT observations were carried 
out at 610 MHz with a bandwidth of 33 MHz. 
The data obtained with the GMRT were channelized 
(512 channels) total intensity data sampled at 
122 $\mu$s. These data were further analyzed 
offline using the SIGPROC pulsar analysis software.   

\section{Analysis and Results}
\label{analysis}

\subsection{Single pulse sequences and long nulls}
\label{analsp}

\begin{figure*}
\includegraphics[width=1.0\textwidth]{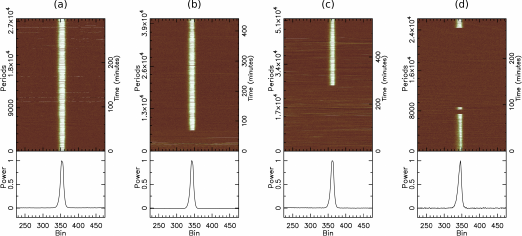}
\caption{Single pulse plots of PSR B1706$-$16 for four 
different observations. The top panel in each figure is 
the single pulse sequence and the bottom panel is the 
integrated profile. The ordinate on the left-hand axis 
in each pulse sequence plot denotes the pulse number, 
whereas the ordinate on the right-hand axis gives 
time in minutes. Figure (a) shows the typical single pulses 
observed with out any significant long nulls. 
Figures (b), (c) and (d) shows the long nulls observed 
at three different epochs.} 
\label{singlepulses}
\end{figure*}

The dedispersed data from the ORT observations 
was folded using the predictors generated with 
the TEMPO2\footnote{http://tempo2.sourceforge.net/} 
software to produce the single pulse sequences as 
shown in the Figure \ref{singlepulses}. Figure 
\ref{singlepulses}(a) shows the typical sequence 
obtained in 11 of the 15 long observing sessions. 
The pulsar is usually in an active state with 
short nulls of null duration not more than 150 
periods, which are easy to identify as the 
single pulses can be seen with high signal 
to noise ratio (S/N). Figure \ref{singlepulses}(b), (c) and (d) 
show 3 out of 4 observing sessions, where long nulls 
were observed, with each long null lasting between 
1 hour to 4.5 hours. This nulling behaviour is 
rare, with the pulsar exhibiting two different 
phases, an Active Phase (AP), with pulsed 
emission seen in most periods, interspersed by short nulls, 
and an Inactive Phase (IP), where it switches 
off for few hours with no pulsations at all,
similar to other intermediate nulling pulsars 
\citep{Young2015}. However, unlike the other 
intermediate nullers, the pulsar is in AP most 
of the time and rarely switches to the IP.

The nulling analysis was performed using the methods  
devised by \cite{Ritchings1976,gjk12}. These methods, used for estimating 
the NF and identifying null and burst pulses are briefly described below 
\citep[For details see][]{gjk12}. We visually identified two 
windows of equal width in the average profile, namely, a window with 
phase bins, where the pulsed emission is present (on-pulse window) 
and another window, away from the pulsed emission (off-pulse window). 
Two sequences of energies, integrated over on-pulse and off-pulse 
windows for each pulse, were formed after normalizing with the 
mean pulse energy. The NF, which represents the percentage of 
pulses with on-pulse energy distribution similar to that of 
off-pulse energy, was estimated from the distributions  
of on-pulse and off-pulse energies 
\citep[See][for details]{gjk12}. 
A threshold energy separating the zero energy excess 
in the on-pulse distribution was used to separate null and burst 
pulses. An overlap in the peak 
of zero energy excess and the burst energy in the on-pulse distribution 
can lead to a mis-identification of nulled (burst) pulses.

A total of approximately 100 hours of data were 
obtained on B1706$-$16 in 15 separate long 
observing sessions. The NF was estimated for 
each observation and is shown in  Figure \ref{nulling}. 
The bars in the plot represent the duration of 
observations indicated in hours on the right 
side of the plot. The dotted line represents 
the variation of the NF for each observation 
and the solid line represents the cumulative NF 
calculated for the total duration of the observations. 

NF varies from 15 \% to 70 \% between observations. 
This variation is mainly due to presence of long nulls 
in some observations. The four long nulls detected 
during our observations are listed in the 
Table \ref{nullvalues}. The relatively large NF for 
the other observations is due to the presence of 
several 5 to 20 minute nulls during the AP, 
where the NF is estimated to be 15$\pm$2 \%. 
The cumulative NF, considering nulls in AP as 
well as IP from all observations, is calculated 
to be 31$\pm$2 \%. It is to be noted that the 
pulsar rarely switches to the IP, which makes 
it difficult to get an accurate NF from short 
observations. Indeed, the nulling fraction was never reported 
in the previous studies of this pulsar.

\begin{table}
 \caption{The four epochs, where long nulls were 
observed in PSR B1706$-$16. In three of these epochs, 
the null was not bounded by burst pulses on both 
sides. Hence, only a lower limit on the null 
duration is listed}
 \label{nullvalues}
 \centering
 \begin{tabular}{|l|c|r|}
  \hline
  MJD of              & Length of null& Duration of null\\
  observations    & (Rotations)   &   (hours) \\
  \hline
  56515.5           & $\ge$ 6393  & $\ge$1.16 \\
  56519.4           & $\ge$ 6194  & $\ge$1.12 \\
  56520.5           & 15825       & 2.87 \\
  56535.5           & $\ge$ 25850 & $\ge$4.68 \\
  \hline
 \end{tabular}
\end{table}

\begin{figure}
\centering
\includegraphics[width=\columnwidth]{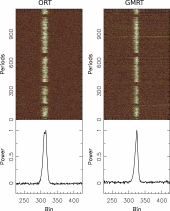}
\caption{Single pulse sequences, observed during the 
simultaneous observations of PSR B1706$-$16 using the 
ORT at 326.5 MHZ and the GMRT at 610 MHz}
\label{simul}
\end{figure}

\begin{figure}
\centering
\includegraphics[width=1\columnwidth]{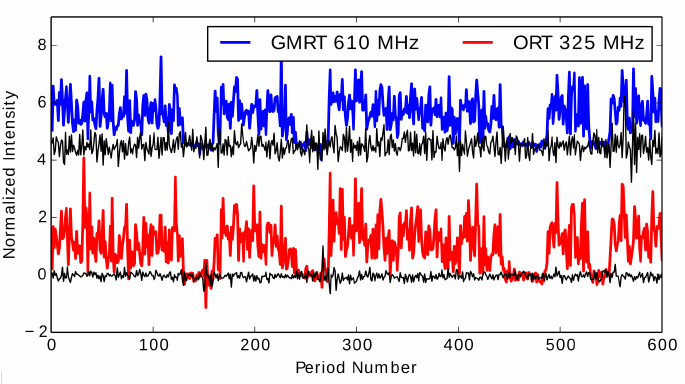}
\caption{Normalized energy in the on-pulse window 
during 325/610 simultaneous observations shown 
in the Figure \ref{simul}. The on-pulse energy at 610  
MHz is shown by the blue solid lines, whereas the red lines 
show the on-pulse energy at 326.5 MHz. The energy integrated 
over off-pulse window is also shown by solid black line for 
comparison and identification of nulls.}
\label{energy}
\end{figure}

\subsection{Is nulling broadband in PSR B1706$-$16?}
\label{bbnull}

The single pulses in  PSR B1706$-$06 were observed 
simultaneously with the ORT at 326.5 MHz and the GMRT 
at 610.0 MHz for a duration of 1 hour (Figure \ref{simul}). 
A short stretch of these observations, shown in this plot, 
clearly shows the burst and null pulses. All the nulls, 
including those lasting a single period 
(not seen in the figure), are observed to be simultaneous.
Likewise, the transition from the burst emission to 
null (and vice versa) is also observed  
to be simultaneous at both the frequencies.  
Figure \ref{energy} shows the on-pulse energy sequence 
for both the observations for a stretch of 500 periods. 
The sharp dips in the on-pulse energy in this figure 
represent the nulls. Again, it is evident that the 
nulls occur at both frequencies simultaneously. Results from a more 
quantitative analysis confirming this conclusion 
are presented below.

\begin{figure}
\centering
\includegraphics[width=\columnwidth]{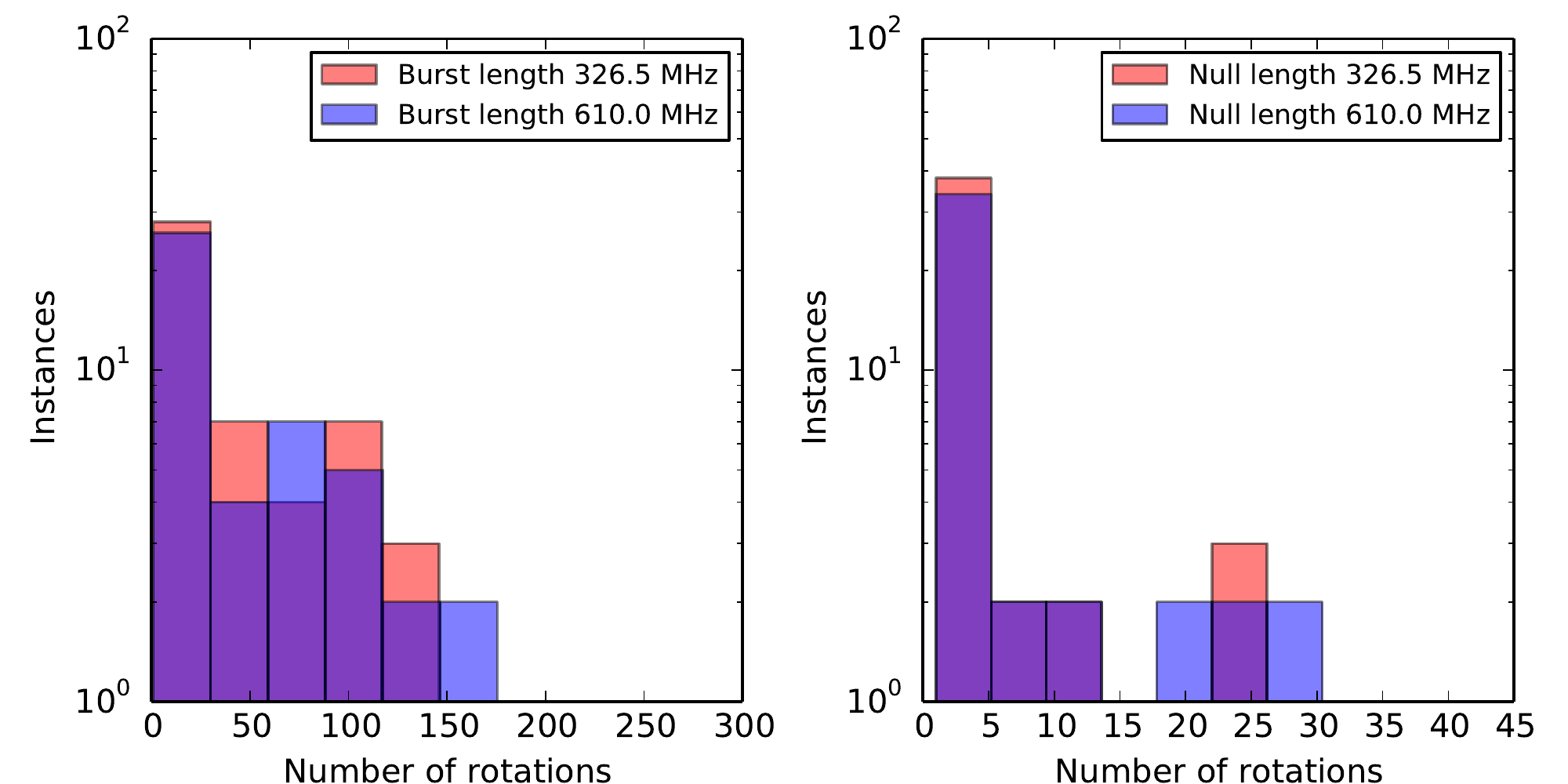}
\caption{Null and burst length histograms of the simultaneous observations.}
\label{nullbursthist}
\end{figure}

Null and burst pulses were identified after 
a careful visual examination of single pulse sequences 
at both the frequencies. A one-bit sequence, representing 
the null and burst pulses (0 and 1 respectively), was derived  
from both the observations for  2844 pulses, excluding 
all the periods affected by radio frequency interference. 
These one-bit sequences 
for both the  observations of the pulsar were compared 
using contingency table analysis, where a 2 X 2 table is 
formed from the two one-bit sequences, giving the statistics 
of nulls and bursts in the two simultaneous one-bit sequences 
(Table \ref{stats}.) In this analysis, 
the strength of correlation is computed using a $\phi$-test 
and uncertainty tests \citep{press}. The estimated Cramer-V 
was 0.97 indicating a high significance of correlation, 
while uncertainty coefficient was estimated to be 0.9,  
consistent with simultaneity of nulling pattern across 
frequencies.  Moreover, 
the NF calculated for the ORT and 
the GMRT observations are 14$\pm$3 \% and 12$\pm$1 \% 
respectively. During the simultaneous observations, 
no long null was observed. The null and burst length 
histograms at both the frequencies 
(Figure \ref{nullbursthist}) were compared using a 
Kolmogorov-Smirnov test, which rejected, at a high 
significance ($\ge$96.9\%), the hypothesis that 
these two distributions are different. Clearly, 
nulling is simultaneous at both frequencies, similar 
to broadband behaviour of nulling demonstrated in 
a previous studies \citep{Gajjar2014,njm+17}, increasing 
the sample of such pulsars.   

\begin{table}
 \caption{Statistics of null and burst pulses in the 
simultaneous 326.5  and 610 MHz observations of PSR B1706$-$16.} 
 \label{stats}
 \centering
 \begin{tabular}{|l|c|c|}
  \hline
                & Null @ 325 MHz& Burst @ 325 MHz\\
  \hline
  Null @ 610 MHz  & 332  &   2 \\
  \hline
  Burst @ 610 MHz & 14  &  2496 \\
  \hline
 \end{tabular}
\end{table}

\begin{figure*}
\includegraphics[width=\textwidth]{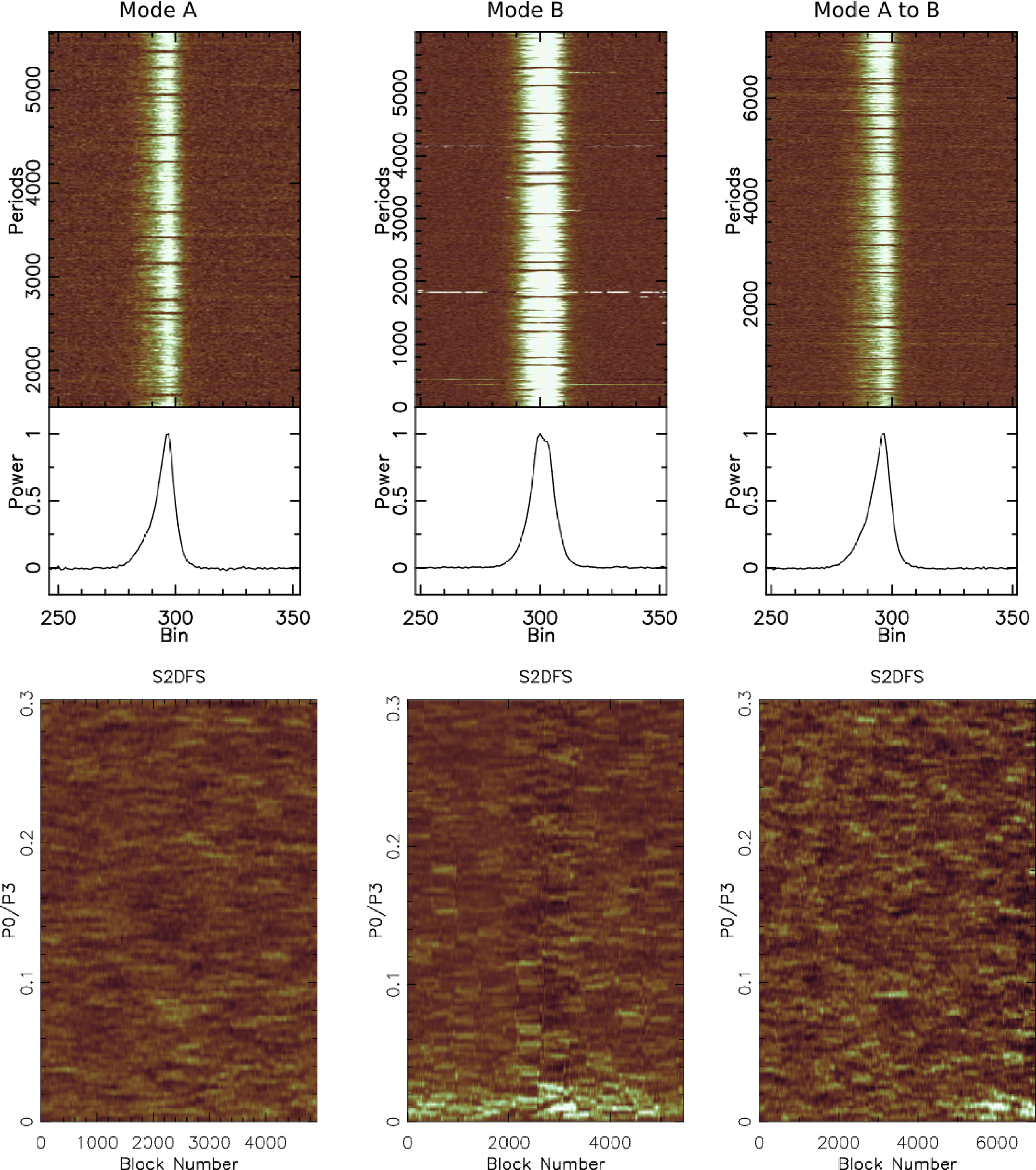}
\caption{Profile modes of PSR B1706$-$16. Top left plot 
shows the profile in mode A and corresponding single 
pulse sequence, which is boxcar averaged with a window 
size of 20 pulses.  The corresponding S2DFS plot 
is shown in bottom. Top middle plot is profile of 
mode B with corresponding boxcar averaged pulse sequence 
and corresponding S2DFS plots at the bottom. The right top 
plot shows an observed mode change in the pulsar. The pulsar 
is in mode A in the first 5500 pulses and switches to mode 
B beyond. This manifests as a drifting signature appearing 
after 5500 pulses in the corresponding S2DFS 
plots in the bottom. }
\label{modes}
\end{figure*}

\subsection{Average profile of PSR B1706$-$16}
\label{avprofanal}

The 610 MHz GMRT observations of PSR B1706$-$16 show 
a single component (see Figure \ref{simul}), which 
according to the EPN pulsar data 
base\footnote{http://www.jb.man.ac.uk/research/pulsar/Resources/epn/} 
shows a steep linear polarization position angle swing 
and a circular polarization sense reversal at the center 
of the component. Thus, this component appears to be 
a core component \citep{Rankin1983}. The corresponding 
profile at 326.5 MHz in the  simultaneous ORT observations 
shows two components. This extra component is absent from 
archival profile in EPN pulsar data base, which suggests 
that we may have observed a profile mode change in our 
simultaneous ORT observations. The EPN archival profile at 
408 MHz shows circular polarization sense reversal 
towards the trailing dominant component. Hence, it 
appears that the extra leading component is a conal 
component. Subpulse drift, manifested as 
a fluctuation periodicity, is expected in this 
leading component as this is usually seen in 
conal components \citep{Rankin1983}. Indeed, we see 
evidence for this, as discussed in Section \ref{profmodanal}, 
alongwith changes in average profile.

\subsection{Profile modes of PSR B1706$-$16}
\label{profmodanal}

As mentioned above, this pulsar seems to exhibit two 
different profile modes at 325 MHz. In mode A, a single 
component profile is seen, while a two component profile 
is observed in mode B (Figure \ref{modes}). One way to 
check if the two profiles are not similar is to treat them 
as histograms and perform a Kolmogorov-Smirnov shape comparison test 
described in \cite{p08}. Results are shown in Table \ref{kstest}, 
which imply that the profiles for the two modes are distinct from each 
other. The two modes are characterized by distinct 
fluctuation properties. As can be seen in Figure 
\ref{modes}, mode B is accompanied by subpulse 
drifting, which is associated with the leading component 
in mode B, while no significant drift feature is detected 
in mode A  (Figure \ref{modes}). The rightmost plots in 
Figure \ref{modes} shows such Sliding window two 
dimensional fluctuation spectra \citep[S2DFS,][]{ssw,njm+17} 
plot, where this transition from mode A to mode B is 
clearly visible at about 5500 periods.

\begin{table}
 \caption{Results of Kolmogorov-Smirnov shape comparison test 
\citep{p08} between the profiles shown in Figure \ref{modes}. 
A significance close to 1 indicates that the two profiles are similar 
and vice-versa}

 \label{kstest}
 \centering
 \begin{tabular}{|c|l|c|}
  \hline
  No              & profiles & Significance \\
  \hline
  1           & mode A and mode B  &   0.13 \\
  2           & before null and after null &   0.23\\
  3           & mode A and before null     &   0.99\\
  4          &  mode A and after null &  0.34\\
  5           & mode B and before null     &   0.25\\
  6          &  mode B and after null &  0.94 \\
  \hline
 \end{tabular}
\end{table}

It should be noted that the ORT is a single polarization 
instrument. If a pulsar is highly polarized and has 
small rotation measure (RM), the profile shape can change 
due to rotation of polarization angle (PA) with respect 
to the telescope feed. PSR B1706$-$16 has a small degree 
of linear polarization and circular polarization 
[11 and 6 \% \citep{gl98}]. It has a small 
RM of $- 1.3~rad/m^2$ \citep{hl87} resulting in a 
swing in polarization angle across 16 MHz band of 
about 6 degrees. Thus, the effect of change in PA 
cannot produce as significant profile change as seen 
in Figure \ref{modes} as most of the emission is 
unpolarized. Moreover, it is evident from S2DFS in 
Figure \ref{modes} that the fluctuation properties of 
profile also change when a profile mode change 
takes place with variable but clear drifting 
seen in the mode B. Thus, the profile change 
between the two modes in PSR B1706$-$16 is not due to 
instrumental effects, but is real. The two profile 
modes accompanied by changes in drift mode for 
this pulsar are being reported for the first 
time in this paper.  

\begin{figure*}
\centering
\includegraphics[width=\textwidth]{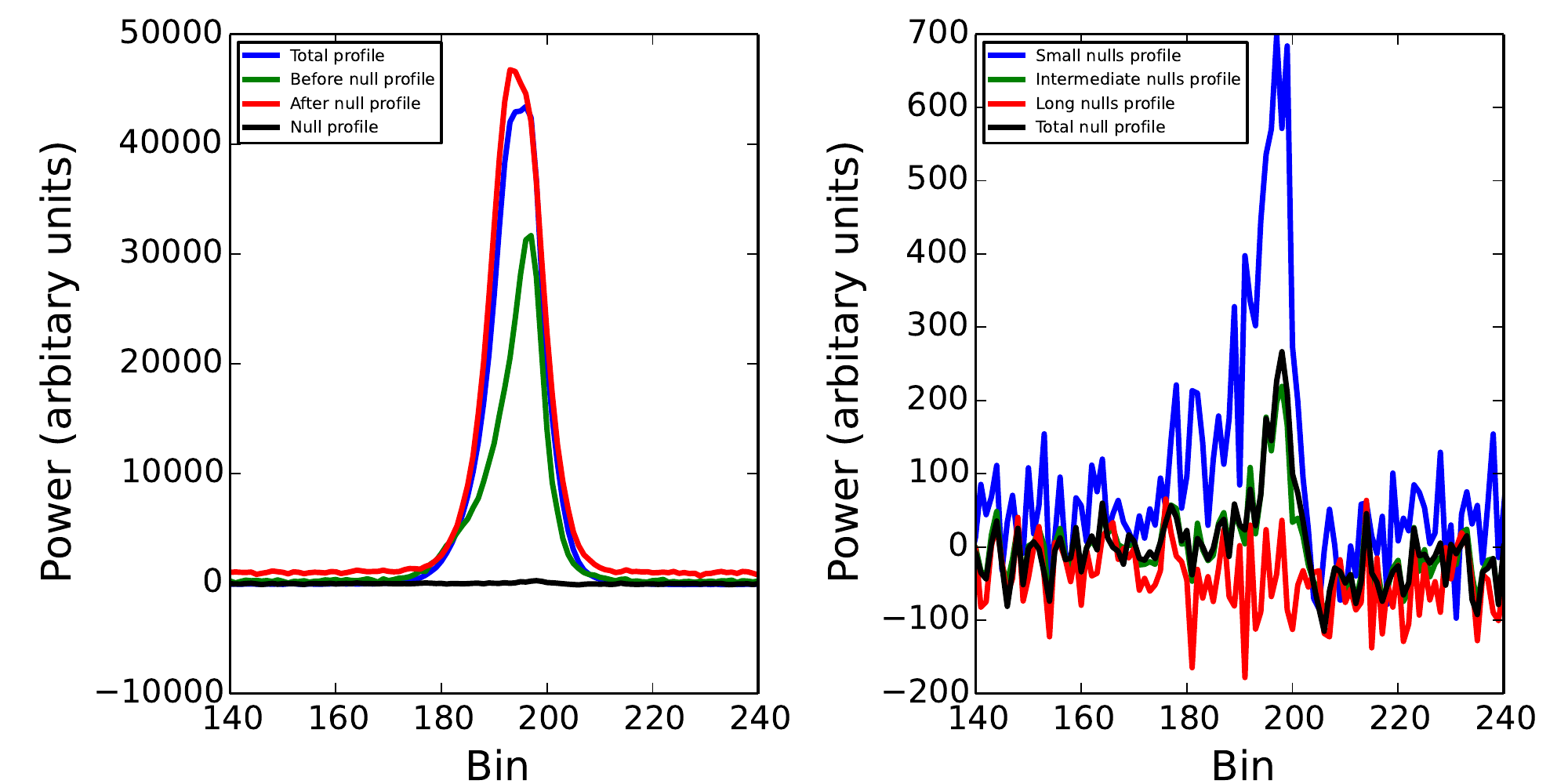}
\caption{Plot on the left shows profiles for 4 different 
sections of data, 1) Average profile after integrating 
all the observations (blue), 2) Average profile after 
integrating 10 pulses just before every null for  15 
long observations with about 7500 nulls (green), 3) 
Average profile after integrating 10 pulses just 
after every null for the 15 long observations (red), 
4) Average null profile for all the nulled pulses  (black). 
Plot on right shows average null profiles for nulls with 
different null duration 1) null profile obtained 
after averaging nulls (black). 2) Null profile obtained after 
averaging all nulls which are less that 10 periods (blue). 
3) Null profile obtained after averaging all the intermediate nulls 
($>$ 10 periods) shown in green. 4) Null profile obtained after 
averaging the 4 long nulls in Table \ref{nullvalues} (red).}
\label{on_off_profiles}
\end{figure*}

\subsection{Average profile before and after null and during nulls}
\label{banullprofanal}

The visual examination of the emission from PSR B1706$-$16 
before a typical null suggests that the emission seems 
to diminish gradually (Figure \ref{energy}). The mean 
pulse intensity just before the null is less than that 
for the full data suggesting that the pulsar switches 
off to the null state gradually. This behaviour appears 
similar to PSR J1752+2359, where differences in profiles 
between last pulse before a null and the one after the null 
were reported by \cite{gjw14}. Hence, we investigated the 
emission before and after nulls. We selected 10 pulses 
before and after every null. The left plot in Figure 
\ref{on_off_profiles} shows the profile just after 
the null (red) and before the null (green) 
along with the total average profile (blue) and 
the null profile (cyan) for all 15 observations. 
This behaviour is observed in all individual data 
sets of PSR B1706$-$16. The pulses just after the 
null were observed to be of much higher intensity 
than those observed on an average in this pulsar. 
Moreover, the before null profile seems to be similar 
to mode A and profile just after the null is similar to 
mode B. An analysis using K-S shape comparison  test also confirms 
this conclusion (See Table \ref{kstest}).

Nulls are usually defined as pulses with no detectable 
emission. However, this does not seem to be strictly 
true for all nulls in PSR B1706$-$16. The right plot in 
the Figure \ref{on_off_profiles} shows the average 
profiles for nulls with various null duration 
using all observations. The average null profile 
for all nulls shows a weak pulse, which is probably 
due to nulls with duration less than one hour (blue 
and green). On the other hand, no emission is seen 
for long nulls ($>$ 1 hour). The ratio of total 
energy in average profile to null profile is 
about $\sim$ 1420. To the best of our knowledge, 
this is the largest drop in the pulsed emission 
during the nulls ever reported 
[see \cite {vj97, gjk12, Gajjar2014}].

\begin{figure*}
\centering
\includegraphics[width=1\textwidth]{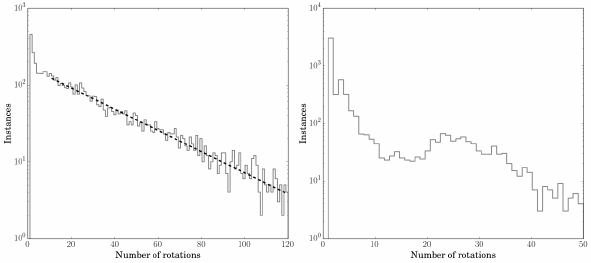}
\caption{Burst length (left plot) and null length (right plot) 
         distributions.}
\label{on_off_length}
\end{figure*}

\subsection{Distribution of burst and null duration}
\label{bndistanal}

\subsubsection{Burst duration distribution}
\label{bldistanal}

The distribution of the duration for which the pulsar 
is in the on state (burst) is shown in Figure 
\ref{on_off_length}. This distribution is exponential 
(black line fit), indicating that the pulsar does not 
have memory about the burst duration after the 
previous null. The slight excess of the short 
bursts may be due to mis-identification of a few 
short nulls as bursts. As discussed in Section 
\ref{analsp}, such mis-identification results from the 
overlap between the zero excess and the burst part of 
on-pulse energy distributions.    

\subsubsection{Null duration distribution}
\label{nldistanal}

The distribution of the null duration of the pulsar,  
shown in the right plot of the Figure 
\ref{on_off_length}, is bimodal. 
Short nulls of few rotations 
significantly out number the longer nulls 
extending to tens of rotations. This is 
partially due to mis-identification of weak low S/N 
burst pulses as nulls, which could 
also be the most probable explanation for the 
weak profile seen after averaging all nulls of duration less 
than 10 periods  
(Figure \ref{on_off_profiles}).  
However, very few pulses are likely to be 
mislabeled with most of these being single period 
nulls. Thus, the bi-modality appears to be genuine. 
A similar behavior was reported for PSR B0031$-$07 
\citep{Vivekanand1995}


\section{Discussion}
\label{discussion}

Our observations show that PSR B1706$-$16 has a NF of 
15$\pm$2\% during the AP and shows sporadically 
very long nulls. The longest null duration 
identified is at least 4.68 hours. Its overall 
NF is 31$\pm$2\%. Simultaneous two frequency 
observations show that nulling in this pulsar 
is broadband. We have, for the first time, 
identified two profile modes in this pulsar, 
which are accompanied by difference in the 
fluctuation properties of single pulses. While the 
emission during the long nulls drops by a factor of 
1420, weak emission is seen in an average profile 
of short nulls. We also report for the first 
time a change in profile before and after nulls, 
with the profile before the null similar to mode 
A profile and that after a null similar to mode B. 
Finally, a bimodal distribution for null length is 
reported.

This interesting nulling behaviour is reported for 
the first time for this pulsar. The variability in 
NF from one observing session to other (Figure 
\ref{nulling}) implies that NF, at best, provides a 
qualitative description of nulling and is not an appropriate parameter 
to  look for correlation with other pulsar parameters 
as was done in some previous studies [e.g. see  \cite{Biggs1992}]. 
Nulling is likely to be better characterized by 
the nulling pattern or null length distribution, as 
was also shown by \cite{gjk12}. This motivates 
longer than usual 1 hour observations to determine 
these distributions before a comparison with pulse 
period, magnetic field or spin-down energy loss 
can be done. 


The simultaneity of nulling at 326.5 MHz and  610 MHz, seen in 
PSR B1706$-$06, is consistent with the previous multi-frequency 
studies of three  pulsars, where the nulling was reported 
to be broadband \citep{Gajjar2014}. Broadly, there are two 
different explanation for pulse nulling. The first class 
of models invokes intrinsic changes in the magnetospheric 
physics to explain nulling, whereas the second class 
invokes geometry as explained in \cite{Gajjar2014}. Our result 
further strengthens the possibility that intrinsic changes 
are responsible for nulling rather than 
the geometrical effects, such as traverses of the 
line-of-sight through the gaps between the sub 
beams (pseudo-nulls). 

The difference in pulse profile shape  with different 
drift modes has been reported in a couple of previous 
studies in PSR B0031$-$07 and B2319+60 
\citep{wf81,vj97,jos13b}. Similar behaviour in PSR B1706$-$06 
further strengthens the possibility that the profile 
mode changes alongwith the nulls are probably related to 
changes in pulsar magnetosphere. 


Most pulsars show exponential or log-normal distributions 
for their null duration. The notable exceptions are 
PSR B0031$-$07 \citep{Vivekanand1995}, J1717$-$4054 
\citep{Kerr2014}, J1649+2533 and 
B2310+42 \citep{wwr+12}, where bimodal nulling distributions 
have been reported. PSR B1706$-$06 also shows a bimodal 
distribution representing short as well as long nulls. Together 
with the on-state (or more precisely two different on-states 
if the two profile modes are taken into account), this represents 
a multi-state Markov process for such state switching as 
proposed recently \citep{gjk12,c13}, which could arise 
from a combination of modulation of ion and electron 
currents within a range of extreme vacuum and 
force-free state \citep{lst12b,lst12a}.

The non-white distribution in the timing analysis 
of the PSR B1706$-$16 \citep{Baykal1999,Hobbs2010} 
suggests that this pulsar exhibits varying slowdown 
rates ($\dot{\nu}$) with time. If the pulsar undergoes 
changes in slowdown rate during the nulls with 
random null duration, it manifests as timing noise 
in the residuals for the pulsar. We also see 
evidence for such timing noise in our data. Such variable 
$\dot{\nu}$ was attributed to switching between 
distinct magnetospheric states for intermittent pulsars 
\citep{Timokhin2010}. While the intermediate nature of nulls 
in PSR B1706$-$16 does not permit establishing this clearly, 
a higher timing noise is certainly expected in this model.

Classical nullers are known to have off-state 
ranging upto several minutes in contrast to 
intermittent pulsars (which null for several days) and 
Rotating Radio transients (RRATs, which show 
isolated single period bursts separated by several 
periods). Long nulls reported by us place PSR 
B1706$-$16 in the growing class of intermediate 
nullers which lie between the classical nullers 
and intermittent pulsars. The intermediate nullers and 
intermittent pulsars are a useful way to probe the 
effect of magnetospheric changes on pulsar timing 
\citep{lhk+10}. While the latter require a several 
years to study their timing behaviour, intermediate 
nullers provide a tool for such studies in a smaller 
time-scale.

It is useful to compare and contrast the nulling 
behaviour of PSR B1706$-$06 with that of other known 
intermediate nullers. While the NF for PSR B0823+26 in its 
AP is estimated to be 15$\pm$1 \% \citep{Sobey2015}, estimates 
for other intermediate nullers range from 67$\pm$8 \% in PSR J1853+0505 
to 
90$\pm$5 \% in 
PSR J1634$-$5107 \citep{Young2015}. It may 
be noted that these may be overestimates as the NF were  
mostly inferred from non-detection/detection statistics in these 
pulsars. Thus, the NF for PSR B1706$-$06 is much smaller in contrast 
with majority of intermediate nuller. Bimodal null 
distribution was reported for PSR J1717$-$4054 
\citep{Kerr2014}, while it is not well determined 
for other intermediate nuller. Thus, more detailed observations 
of other intermediate nuller are needed to check if this property 
is shared by this class as a whole. PSRs B0823$+$26, 
J1634$-$5107 and PSR J1853$+$0505 all show weak emission 
during longer nulls \citep{Young2015}, whereas  we do not 
detect any weak emission in PSR B1706$-$16 after 
integrating all long nulls, although weak emission is 
seen for shorter nulls in AP. It may be noted that no 
weak emission was detected in the PSR J1717$-$4054 
\citep{Kerr2014,Young2015}. Lastly, high timing noise 
or non-white timing residuals have been noted for 
PSR J1634$-$5107 and PSR J1717$-$4054, while such 
behaviour is not very apparent in PSR J1853+0505 
and B0823+26. The constraints on time scales 
for IP for the latter two pulsars 
are not very stringent at the moment and short 
IP time-scale may explain this difference. 
Longer (8$-$10 hrs) and more frequent 
observations (with a cadence of 1$-$2 days) of an enhanced sample 
of intermediate nullers is therefore motivated to investigate 
a correlation between timing noise and long nulls 
in these pulsars.  Future 
multi-beam telescopes, such as the SKA, MWA and LOFAR, will be very useful 
in such studies as they will not only provide higher 
sensitivity for unambiguous classifications of nulls, 
but also a commensal way of observing with other pulsar 
programs, such as search and timing, due to availability 
of multiple beams.

\section{Conclusions}
\label{conclusion}
This paper presents the results from the nulling analysis of over 100 hours data on the PSR B1706$-$16 observed using the ORT. This pulsar is observed in 15 long observations with the duration of observations varying from 2 to 7.5 hours. It exhibits  long nulls (null duration of $>$ 2 hours) suggesting that it is an intermediate nuller. Typical intermediate nullers have large nulling fractions ($>$ 70 \%) due to the frequent long nulls. However, long nulls are seen infrequently, typically once in a week, in this pulsar making it a unique addition to the intermediate nullers list. The over all nulling fraction is estimated to be $31\pm2$ \%. It shows a bimodal distribution for null duration. The pulsar's integrated profile is observed to switch from one mode to another with different fluctuation properties in the two modes. 

\section*{Acknowledgments}

We thank the staff of the Ooty Radio Telescope and the Giant Meterwave
Radio Telescope for making these observations possible. Both these
telescopes are operated by National Centre for Radio Astrophysics (TIFR).
This work made use of PONDER backend, built with TIFR XII plan grants
12P0714 and 12P0716. We like to thank the anonymous referee for his/her useful
comments and suggestions. BCJ, PKM and MAK acknowledge support for this work
from DST-SERB grant EMR/2015/000515.

\bibliographystyle{mnras}
\bibliography{references} 



\appendix

\bsp	
\label{lastpage}
\end{document}